\documentclass[cits]{PoS}

\usepackage[utf8]{inputenc}
\usepackage{amsmath,overpic}
\usepackage[href]{journals}

\title{Precision Measurement of CP Violation in $D^0\to\pi^+\pi^-$ at CDF}

\ShortTitle{HQL 2010}

\author{Angelo Di Canto
	\thanks{Speaker on behalf of the CDF Collaboration.}
        \thanks{I would like to thank the HQL 2010 organizers for the opportunity of speak at this conference
        and the colleagues of the CDF Collaboration for assisting me while preparing the talk and this document.}\\
        INFN \& University of Pisa, Fermilab\\
        E-mail: \email{angelo.dicanto@pi.infn.it}}

\abstract{We report a preliminary measurement of the CP violating asymmetry in \mbox{$D^0\to\pi^+\pi^-$} using approximately $215^\cdot000$ decays reconstructed in about $5.94$ fb$^{-1}$ of CDF data. We use the strong $D^{\star +}\to D^0\pi^+$ decay (``$D^{\star}$ tag'') to identify the flavor of the charmed meson at production time and exploit CP-conserving strong $c\bar{c}$ pair-production in $p\bar{p}$ collisions. Higher statistic samples of Cabibbo-favored $D^0\to K^-\pi^+$ decays with and without $D^{\star}$ tag are used to highly suppress systematic uncertainties due to detector effects. The result is the world's most precise measurement to date.}

\FullConference{The Xth Nicola Cabibbo International Conference on Heavy Quarks and Leptons,\\
		October 11-15, 2010\\
		Frascati (Rome) Italy}

\newcommand{\Acp}{\ensuremath{A_{\text{CP}}}}
\newcommand{\acp}[1]{\ensuremath{a_{\text{CP}}^{\text{#1}}}}
\newcommand{\Acpraw}{\ensuremath{\Acp^{\text{raw}}}}
\newcommand{\stat}{\ensuremath{\mathit{~(stat.)}}}
\newcommand{\syst}{\ensuremath{\mathit{~(syst.)}}}
\newcommand{\Dbar}{\ensuremath{\overline{D}{}}}

\begin{document}
\graphicspath{{figs/}}

\section{Introduction}
Time integrated CP-violating asymmetries of singly-Cabibbo transitions as $D^0\to\pi^+\pi^-$ and $D^0\to K^+ K^-$ are powerful probes of new physics (NP). Contribution to these decays from ``penguin'' amplitudes are negligible in the Standard Model (SM), but presence of NP particles could enhance the size of CP violation with respect to the SM expectation. Any asymmetry significantly larger than few $0.1\%$, as expected in the CKM hierarchy, may unambiguously indicate new physics contributions \cite{theory}.

We present a high statistic search for CP violation in the $D^0\to\pi^+\pi^-$ decay through the measurement of the time-integrated CP asymmetry:
\begin{equation}\label{eq:acp}
\Acp(\pi^+\pi^-)=\frac{\Gamma(D^0\to\pi^+\pi^-)-\Gamma(\Dbar^0\to\pi^-\pi^+)}{\Gamma(D^0\to\pi^+\pi^-)+\Gamma(\Dbar^0\to\pi^-\pi^+)}
\approx \acp{dir} + \frac{\langle t \rangle}{\tau}\acp{ind}.
\end{equation}
This asymmetry, owing to the slow mixing rate of charm mesons, is to first order the linear combination of a direct, $\acp{dir}$, and an indirect, $\acp{ind}$, term through a coefficient that is the mean proper decay time of $D^0$ candidates, $\langle t \rangle$, in unit of $D^0$ lifetime ($\tau \approx 0.5$ ps). Since the value of $\langle t \rangle$ depends on the specific observed proper time distribution different experiments measure different values of $\Acp(\pi^+\pi^-)$.

The measurement, described with further details in \cite{cdfnote}, has been perfomed on about $5.94$~fb$^{-1}$ of $p\bar{p}$ collisions at $\sqrt{s}=1.96$ TeV recorded by the CDF~II detector at Fermilab's Tevatron collider.

\section{Analysis overview}
We measure the asymmetry using $D^0\to\pi^+\pi^-$ decays from charged $D^\star$ mesons through fits of the $D^0\pi$ mass distributions. The observed asymmetry includes a possible tiny contribution from actual CP violation, diluted in much larger effects from instrumental charge-asymmetries. We exploit a fully data-driven method that uses higher statistic samples of $D^\star$-tagged (indicated with an asterisk) and untagged Cabibbo-favored $D^0\to K^-\pi^+$ decays to correct for all detector effects thus suppressing systematic uncertainties to below the statistical ones. The uncorrected ``raw'' asymmetries \footnote{``Raw'' are the observed asymmetries in signal yields, $$\Acpraw(D^0\to f) = \frac{N_{\text{obs}}(D^0\to f)-N_{\text{obs}}(\Dbar^0\to\bar{f})}{N_{\text{obs}}(D^0\to f)+N_{\text{obs}}(\Dbar^0\to\bar{f})},$$before any correction for instrumental effects has been applied.} in the three samples can be written as a sum of several contributions:
\begin{align*}
\Acpraw(\pi\pi^\star) &= \Acp(\pi\pi) + \delta(\pi_s)^{\pi\pi^\star}\\
\Acpraw(K\pi^\star) &= \Acp(K\pi) + \delta(\pi_s)^{K\pi^\star} + \delta(K\pi)^{K\pi^\star}\\
\Acpraw(K\pi) &= \Acp(K\pi) + \delta(K\pi)^{K\pi},
\end{align*}
where
\begin{itemize}
\item $\Acp(\pi\pi)$ and  $\Acp(K\pi)$ are the actual physical asymmetries; 
\item $\delta(\pi_s)^{\pi\pi^\star}$ and $\delta(\pi_s)^{K\pi^\star}$ are the instrumental asymmetries in reconstructing a positive or negative soft pion associated to a $\pi^+\pi^-$ and a $K^+\pi^-$ (or $K^-\pi^+$) charm decay. This is mainly induced by charge-asymmetric track-reconstruction efficiency at low transverse momentum.
\item $\delta(K\pi)^{K\pi}$ and $\delta(K\pi)^{K\pi^\star}$ are the instrumental asymmetries in reconstructing a $K^+\pi^-$ or a $K^-\pi^+$ charm decay respectively for the untagged and the $D^\star$--tagged case. These are mainly due to the difference in interaction cross-section with matter between positive and negative kaons. Smaller effect are due to charge-curvature asymmetries in track triggering and reconstruction.
\end{itemize}
The physical asymmetry is extracted by subtracting the instrumental effects through the combination
\begin{equation}\label{eq:formula}
\Acp(\pi\pi) = \Acpraw(\pi\pi^\star) - \Acpraw(K\pi^\star) + \Acpraw(K\pi),
\end{equation}
that is valid if kinematics distributions are equal across samples. Any instrumental effect can vary as a function of a number of kinematic variables or environmental conditions in the detector, but if the kinematic distributions of soft pions are consistent in $K\pi^\star$ and $\pi\pi^\star$ samples, and the distributions of $D^0$ decay products are consistent in $K\pi^\star$ and $K\pi$ samples, then $\delta(\pi_s)^{\pi\pi^\star} \approx \delta(\pi_s)^{K\pi^\star}$ and $\delta(K\pi)^{K\pi^\star}\approx \delta(K\pi)^{K\pi}$. This condition was verified in the analysis by inspecting a large set of kinematic distributions and applying small corrections (reweight) when needed.

\section{Measurement}
The trigger selects pair of tracks from oppositely charged particles that are consistent with originating from a secondary decay vertex separated from the beamline, requiring an impact parameter greater than $100~\mu$m. Using these tracks we reconstruct signals consistent with the desired two-body decays ($\pi^+\pi^-$ or $K^-\pi^+$ or $K^+\pi^-$) of a neutral charmed meson ($D^0$ or $\Dbar^0$). To remove most part of non-promptly produced charmed mesons we also require the impact parameter of the $D^0$ candidate not to exceed $100\ \mu$m. Then we associate a low-momentum charged particle to the meson candidate to construct a $D^{\star+}$ (or $D^{\star-}$) candidate. The flavor of the charmed meson is determined from the charge of the pion in the strong \mbox{$D^{\star+}\to D^0\pi^+$} (or \mbox{$D^{\star-}\to \Dbar^0\pi^-$}) decay. Sample-specific mass requirements are used for the two tagged samples: we ask the two-body invariant mass ($M(K\pi)$ for the $D^0\to K\pi$ case and $M(\pi\pi)$, for the $D^0\to\pi\pi$ case) to lie within $24$ MeV/c$^2$ of the nominal $D^0$ mass.

We reconstruct approximately $215^\cdot000$ $D^\star$--tagged $D^0\to\pi^+\pi^-$ decays, 5 million $D^\star$--tagged $D^0\to\pi^+K^-$ decays and 29 million $D^0\to\pi^+K^-$ decays where no tag was required. The much larger statistics of $D^0\to\pi^+K^-$ channels, with respect to the signal sample, is used for correction of instrumental asymmetries and ensures smaller systematic uncertainties than statistical ones on the final result.

\begin{figure}[p]
\centering
\begin{overpic}[height=0.3\textheight,grid=false]{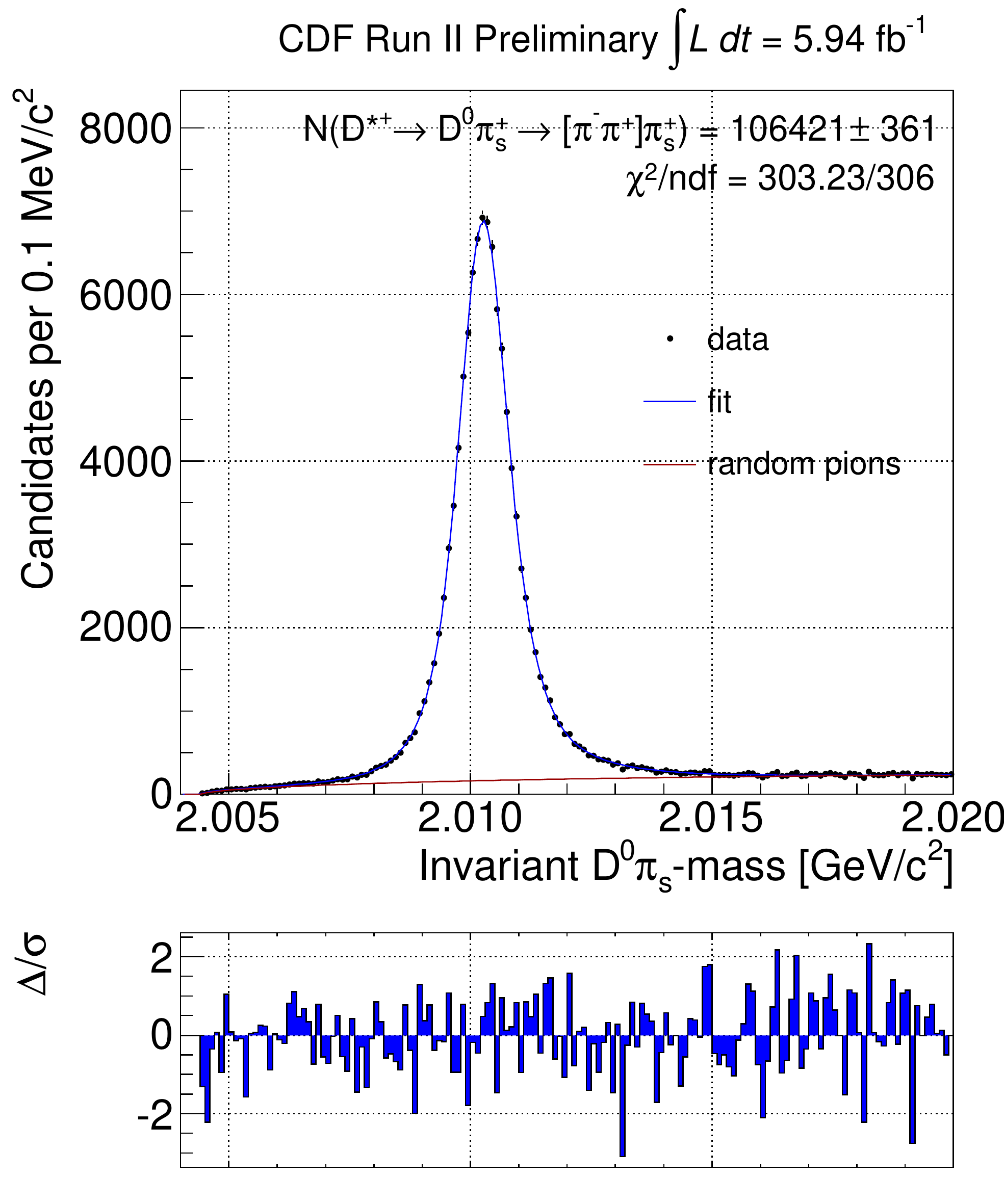}
\put(70,40){\small(a)}
\end{overpic}\hfil
\begin{overpic}[height=0.3\textheight,grid=false]{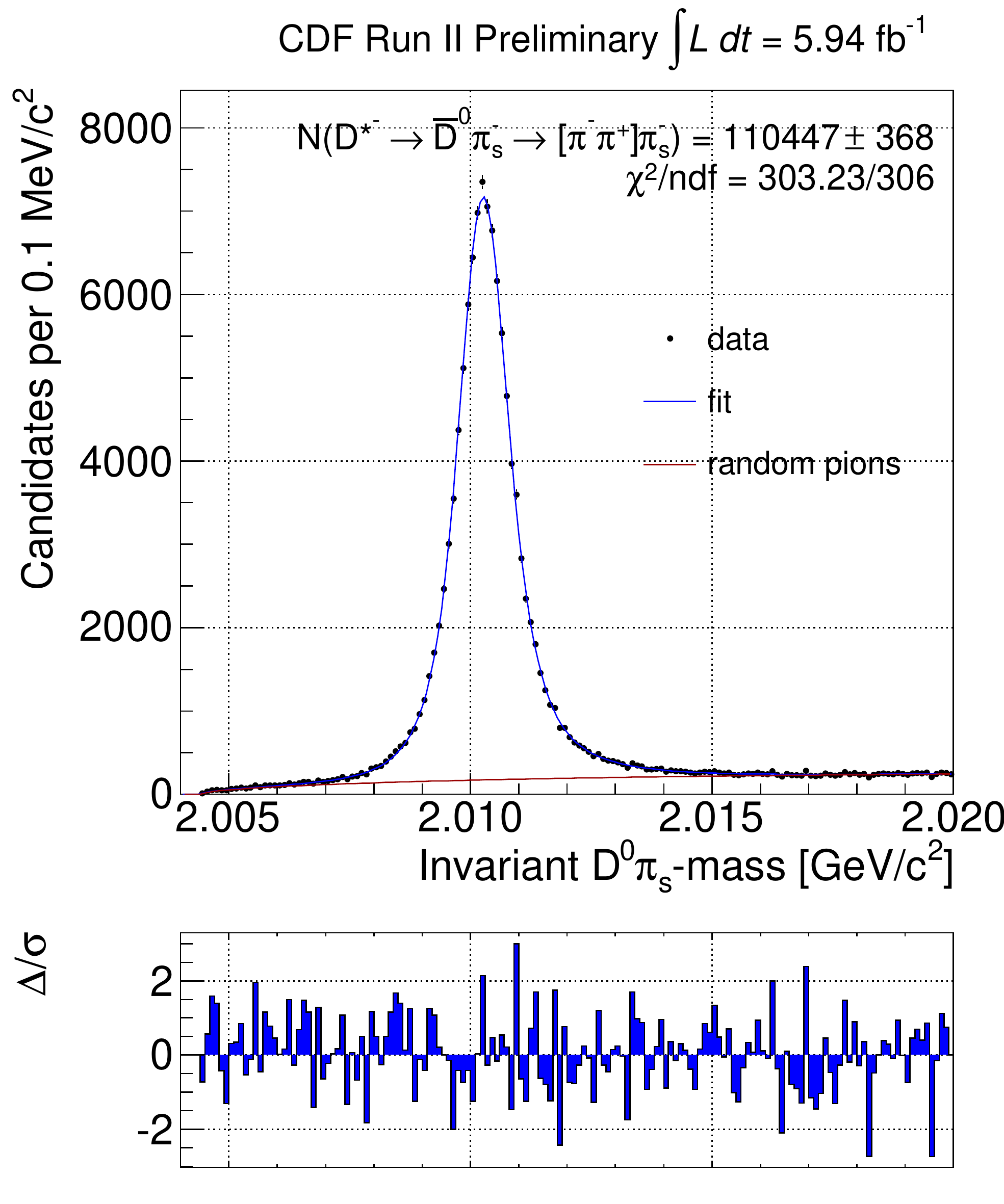}
\put(70,40){\small(b)}
\end{overpic}\\
\begin{overpic}[height=0.3\textheight,grid=false]{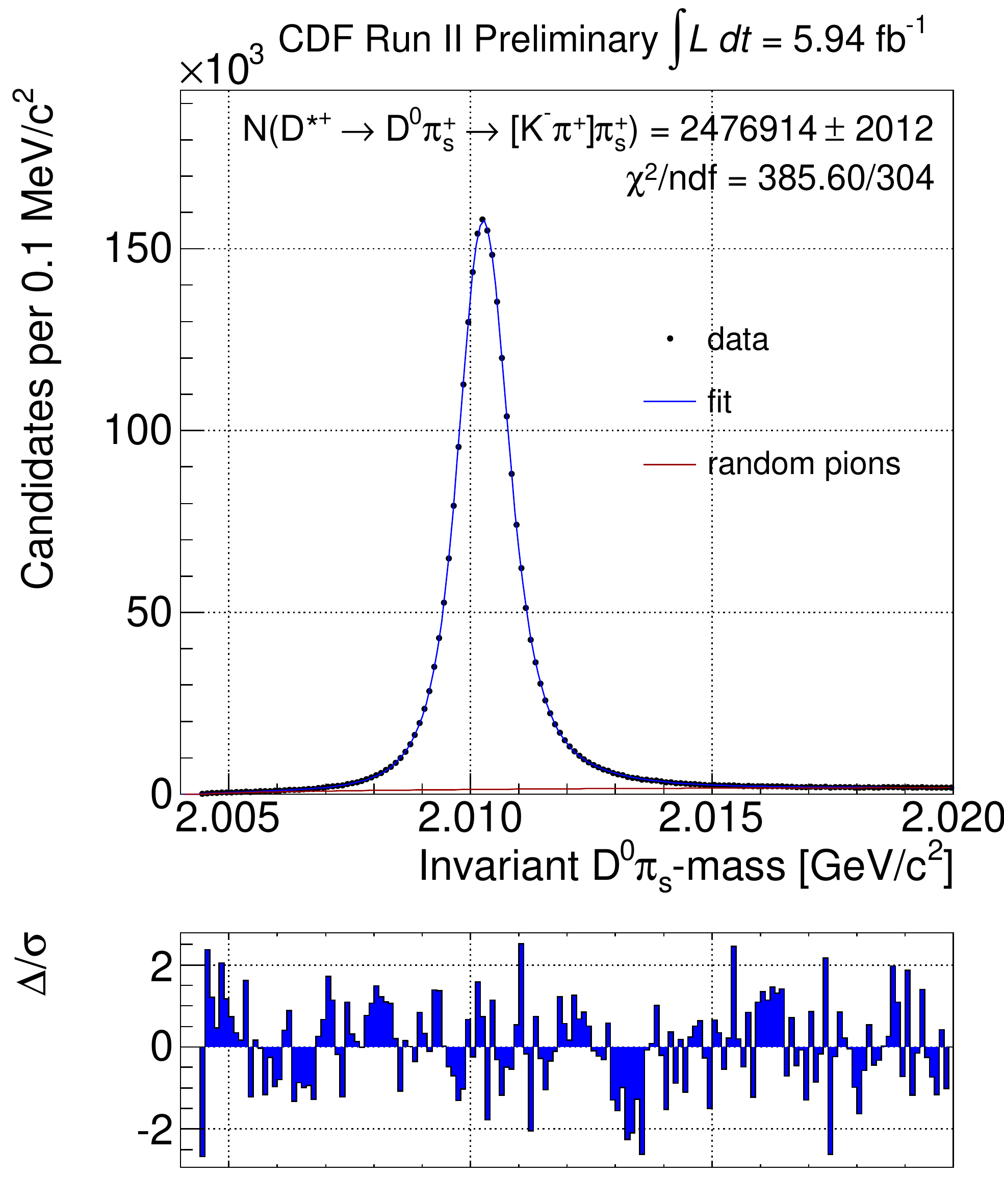}
\put(70,40){\small(c)}
\end{overpic}\hfil
\begin{overpic}[height=0.3\textheight,grid=false]{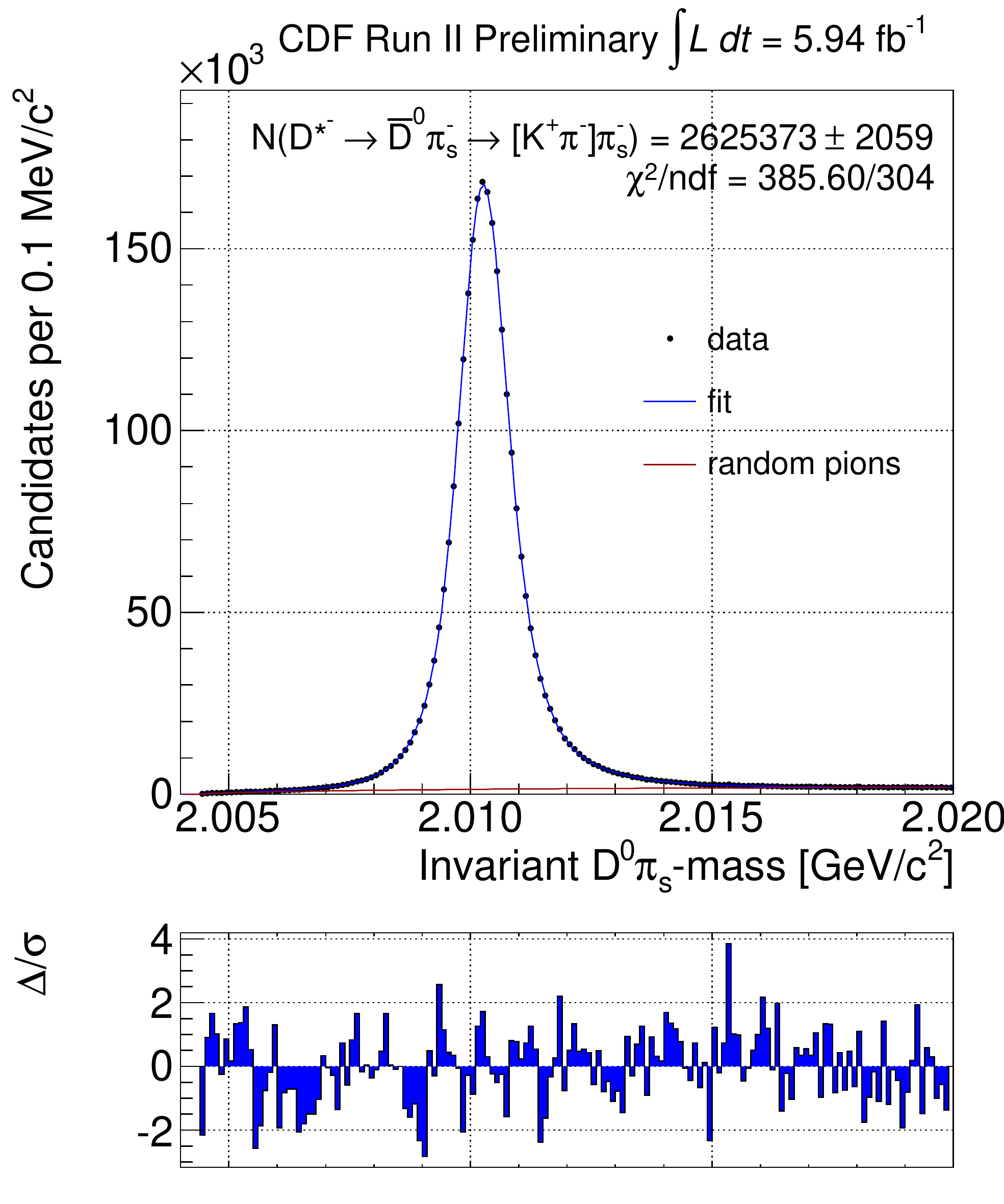}
\put(70,40){\small(d)}
\end{overpic}\\
\begin{overpic}[height=0.3\textheight,grid=false]{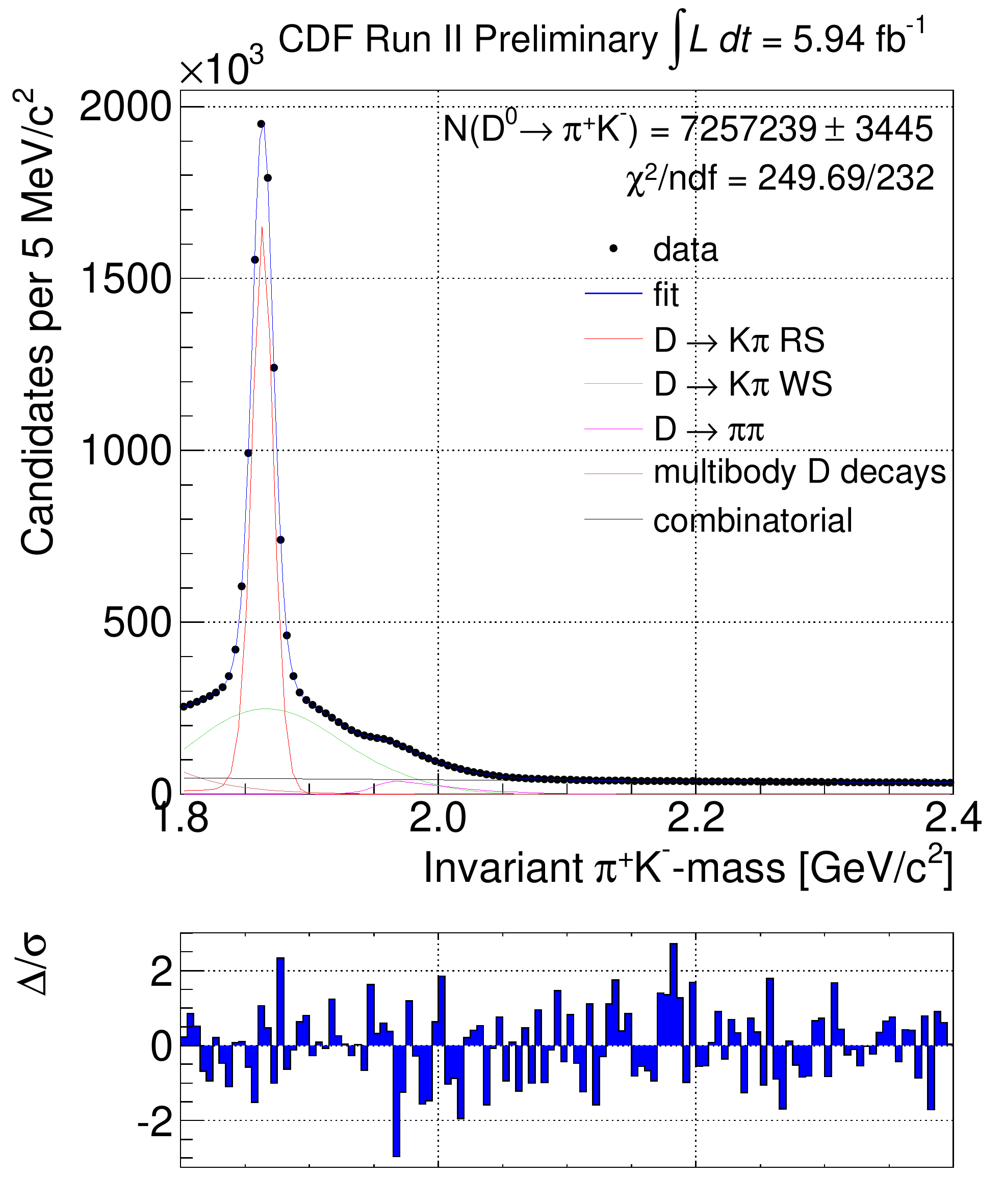}
\put(70,40){\small(e)}
\end{overpic}\hfil
\begin{overpic}[height=0.3\textheight,grid=false]{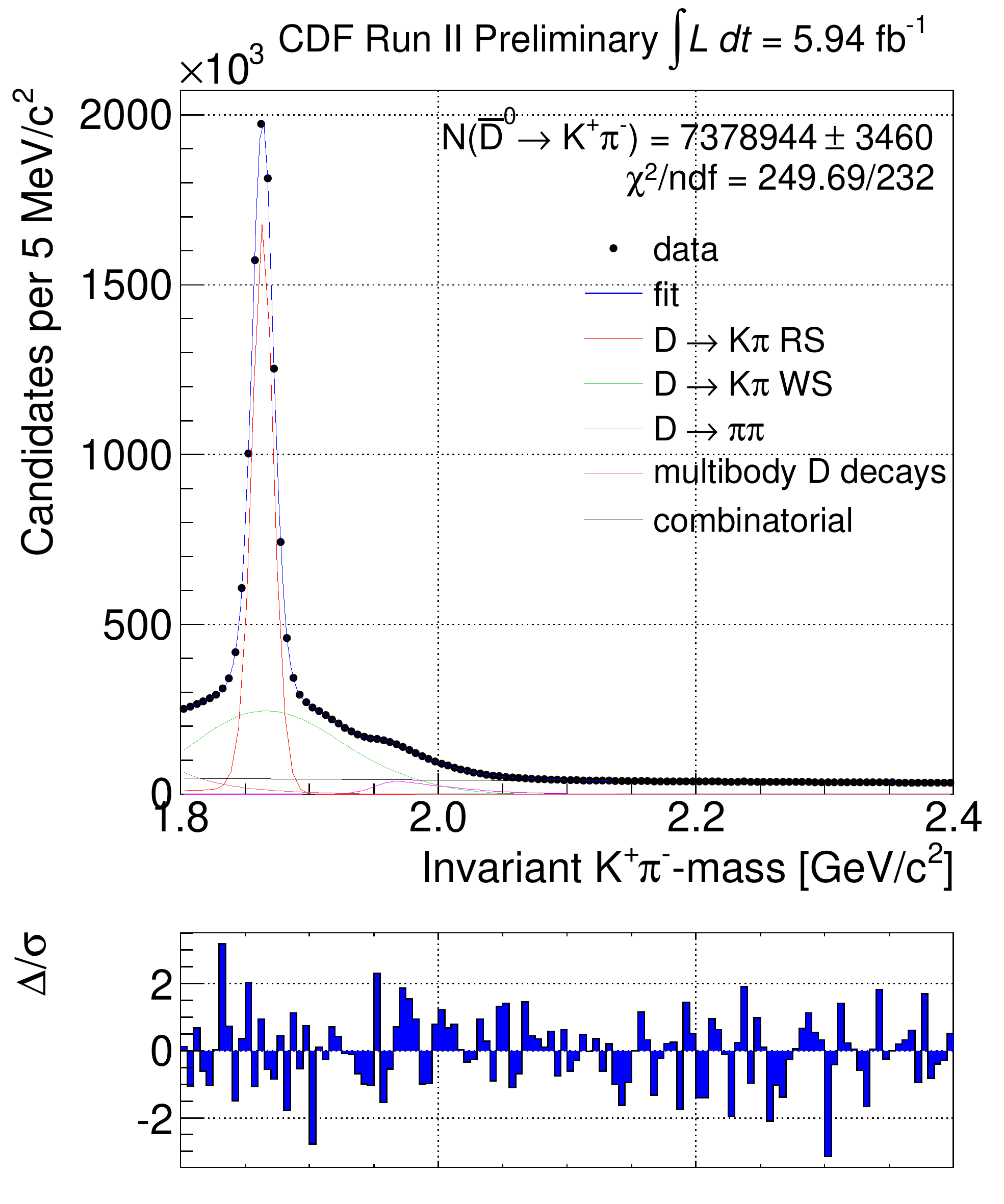}
\put(70,40){\small(f)}
\end{overpic}
\caption{Projections of the combined fit on data for tagged $D^0\to\pi^+\pi^-$ (a)-(b), tagged $D^0\to K^-\pi^+$ (c)-(d) and untagged $D^0\to K^-\pi^+$ (e)-(f) decays. Charm decays on the left and anticharm on the right.}\label{fig:fits}
\end{figure}

We extract independent signal yields for $D^0$ and $\Dbar^0$ candidates without using particle identification in the analysis. In the two $D^\star$-tagged samples this is done using the charge of the soft pion. In the untagged $D^0\to K^-\pi^+$ sample we randomly divided the sample in two indipendent subsamples similar in size. In each subsample we calculate the mass of each candidate with a specific mass assignments: $K^-\pi^+$ in the first subsample and $K^+\pi^-$ in the second one. In one sample the $D^0\to K^-\pi^+$ signal is correctly reconstructed and appears as a narrow peak (about $8$~MeV/c$^2$ wide), overlapping a $\sim10$ times broader peak of the misreconstructed $\Dbar^0\to K^+\pi^-$ component (red and green curves in figs.~\ref{fig:fits}~(e)-(f)). The viceversa applies the other sample. The yield asymmetry is extracted by fitting the number of candidates populating the two narrow peaks.

We determine the yields by performing a binned $\chi^2$ fit to the $D^0\pi_s$-mass ($K\pi$-mass) distribution combining positive and negative decays of both tagged (untagged) samples. The fits projections are shown in fig.~\ref{fig:fits}, the resulting raw asymmetries are: \mbox{$\Acp^{\text{raw}}(\pi\pi^\star)= (-1.86\pm0.23)\%$}, \mbox{$\Acp^{\text{raw}}(K\pi^\star)=(-2.91\pm0.05)\%$}, \mbox{$\Acp^{\text{raw}}(K\pi)= (-0.83\pm0.03)\%$}.

\section{Systematic uncertainties}
The analysis has been tested using Monte Carlo samples simulated with a wide range of physical and detector asymmetries to verify that the cancellation works regardless of the specific configuration. These studies confirm the validity of our approach and provide a quantitative estimate of the systematic errors coming from the basic assumptions in the method. All other systematic uncertainties are evaluated from data. In most cases, this implied varying slightly the shape of the functional forms used in fits (templates), repeating the fit on data, and using the difference between the results of these and the central fit as a systematic uncertainty. A summary of all contributions to the final systematic error is shown in tab.~\ref{tab:syst}.

\begin{table}[t]
\centering
\resizebox{0.6\textwidth}{!}{
\begin{tabular}{|l|c|}
\hline
Source of systematic uncertainty & Variation on $\Acp(\pi\pi)$ \\
\hline
Approximations in the method & $0.009\%$ \\
Beam drag effects & $0.004\%$ \\
Contamination of non-prompt $D^0$s & $0.034\%$ \\
Templates used in fits & $0.010\%$ \\
Templates charge differences & $0.098\%$ \\
Asymmetries from non-subtracted backgrounds & $0.018\%$ \\
Imperfect sample reweighing & $0.0005\%$ \\
\hline
Sum in quadrature & $0.105\%$\\
\hline
\end{tabular}
}
\caption{Summary of systematic uncertainties. Assuming they are independent and summing in quadrature we obtain a total systematic uncertainty on our final $\Acp(\pi^+\pi^-)$ measurement of $0.11\%$.}\label{tab:syst}
\end{table}

\section{Final result and conclusions}
We measure the CP asymmetry in the decay $D^0\to\pi^+\pi^-$ to be
$$\Acp(D^0\to\pi^+\pi^-) = \bigl[+0.22\pm0.24\stat\pm0.11\syst\bigr]\%,$$
which is consistent with CP conservation and also with the SM predictions. 

\begin{figure}[t]
\centering
\begin{overpic}[width=0.33\textwidth,grid=false]{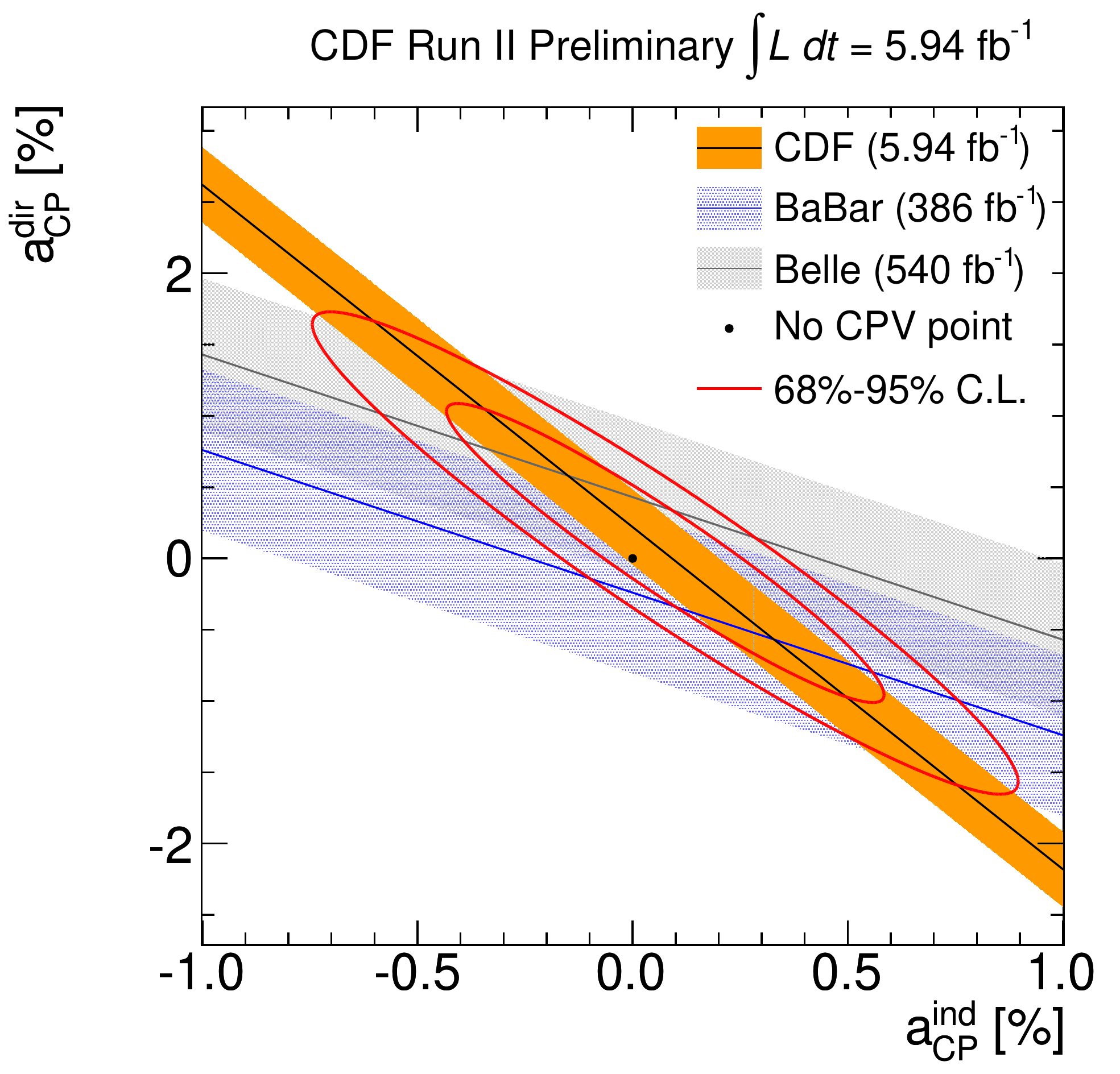}
\put(22,15){\small (a)}
\end{overpic}
\begin{overpic}[width=0.33\textwidth,grid=false]{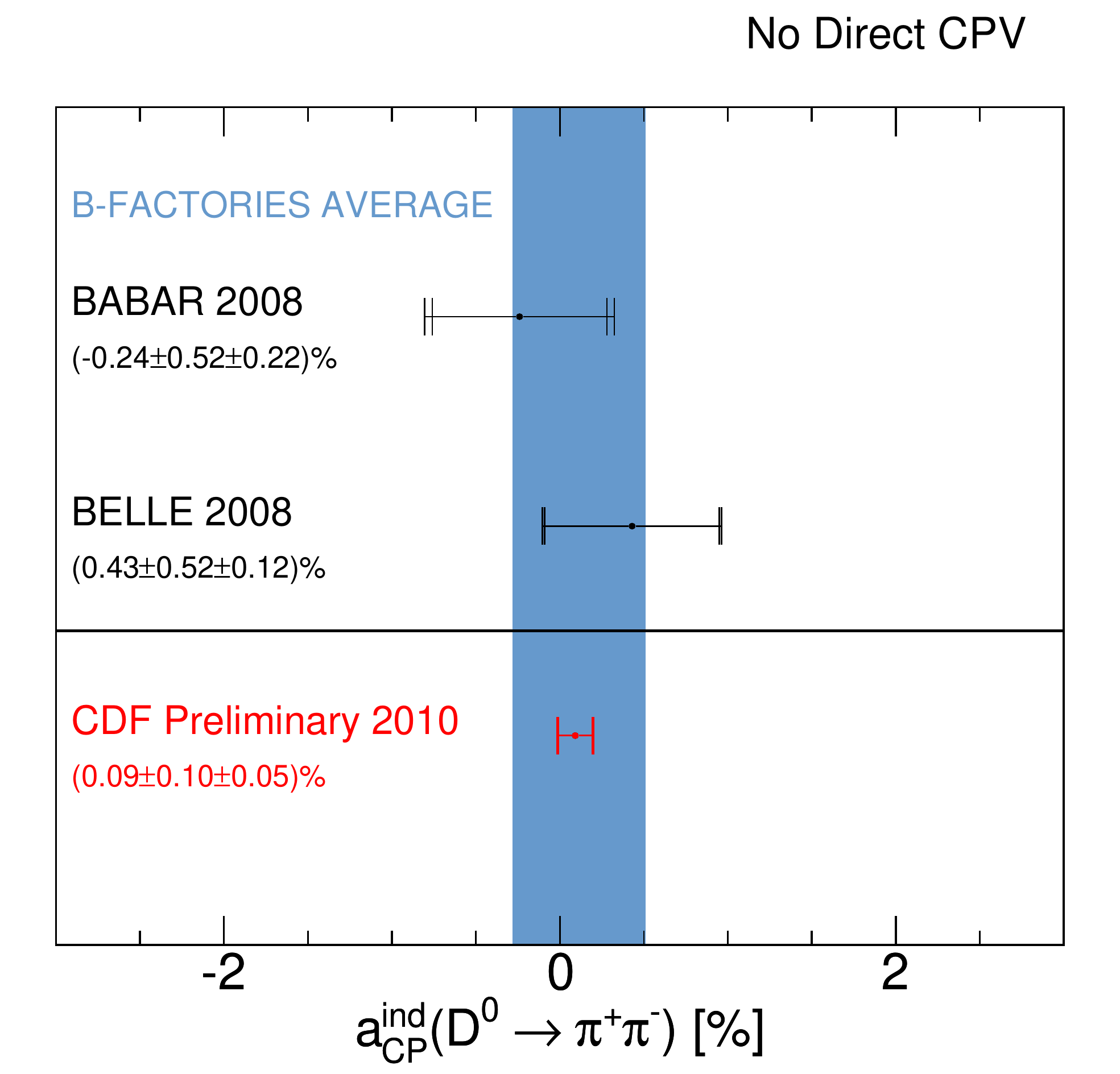}
\put(10,15){\small (b)} 
\end{overpic}\hfil
\begin{overpic}[width=0.33\textwidth,grid=false]{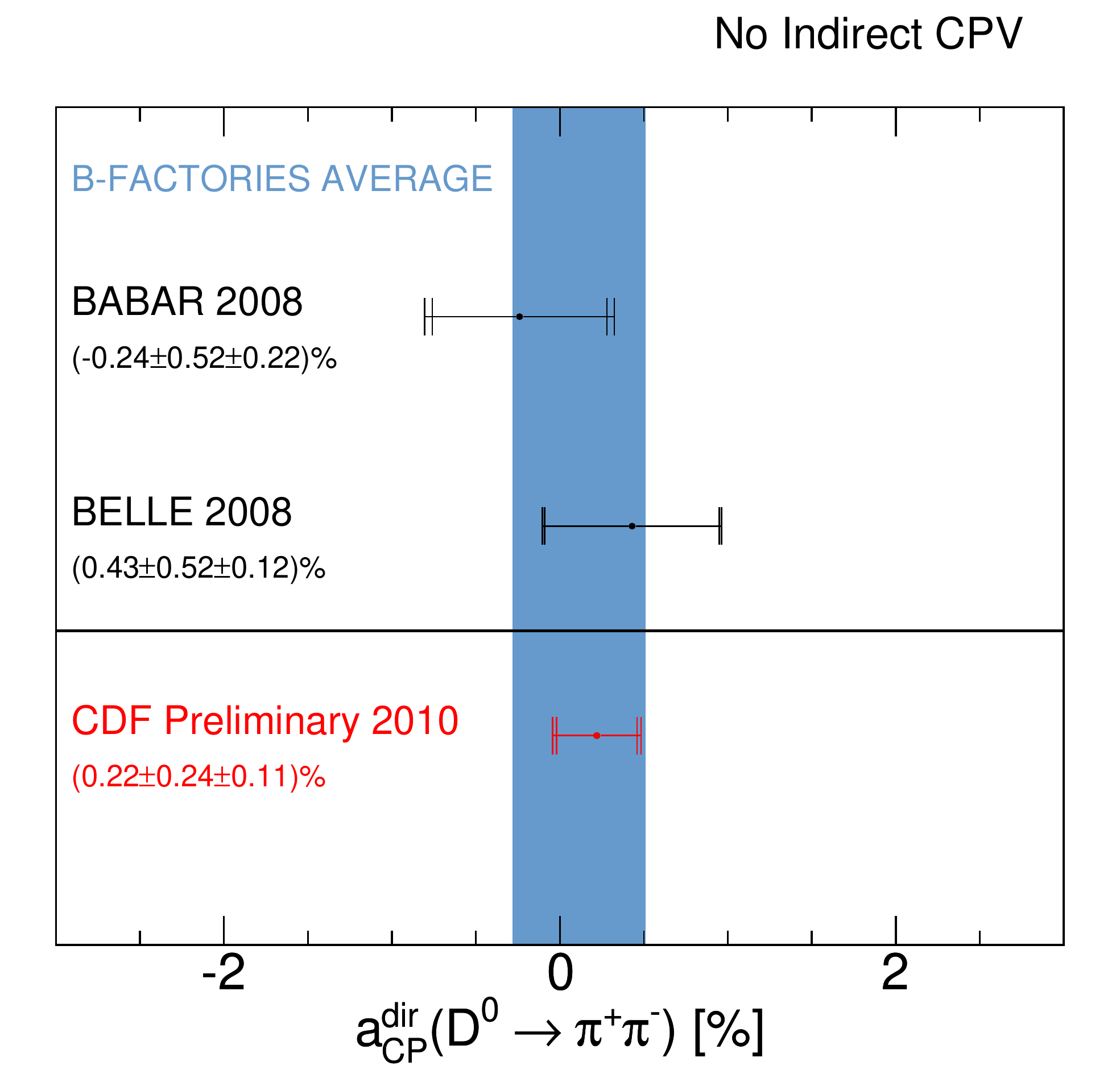}
\put(10,15){\small (c)}
\end{overpic}
\caption{Comparison of our measurement with current best results from B-factories \cite{babelle} in the parameter space $(\acp{ind},\acp{dir})$ (a), assuming no direct (b) or indirect (c) CP violation.}\label{fig:direct_and_indirect}
\end{figure}

As expressed by eq.~\eqref{eq:acp} the $\Acp(\pi^+\pi^-)$ measurement describes a straight line in the plane $(\acp{ind},\acp{dir})$ with angular coefficient given by $\langle t\rangle/\tau$. Because of a threshold on the impact parameter of tracks, imposed at trigger level, our sample of $D^0$ mesons is enriched in higher-valued proper decay time candidates with a mean value of $2.40\pm0.03\ (\mathit{stat.}+\mathit{syst.})$ times the $D^0$ lifetime, as measured from a fit to the proper time distribution. Due to their unbiased acceptance in charm decay time, B-factories' samples have insted $\langle t\rangle=\tau$ \cite{babelle}. Hence, the combination of the three measurements allow to constrain indipendently both $\acp{dir}$ and $\acp{ind}$. Fig.~\ref{fig:direct_and_indirect}~(a) shows such combination: the bands are $1\sigma$ wide and the red curves represent the $68\%$ and $95\%$ CL regions of the combined result assuming Gaussian uncertainties.

Assumimg no direct CP violation in these decays this measurement implies
$$\acp{ind} = \frac{\tau}{\langle t\rangle}\ \Acp(\pi^+\pi^-) = \bigl[+0.09\pm0.10\stat\pm0.05\syst\bigr]\%,$$
that means the range $[-0.124,0.307]\%$ covers $\acp{ind}$ at the 95\% CL. This range is more than five times tighter than the ones obtained using B-factories measurements, as shown in fig.~\ref{fig:direct_and_indirect}~(b).

Conversely, in the assumption of $\acp{ind}=0$, our number is directly comparable to other measurements in different experimental configurations. In this case, fig.~\ref{fig:direct_and_indirect}~(c), our statistical uncertainties are half those from the best B-factories measurements, and systematic uncertainties are also smaller.

We have measured the CP asymmetry in the $D^0\to\pi^+\pi^-$ decay with unprecedented precision, and find a result compatible with zero. An even more precise measurement is expected from the channel $D^0\to K^+K^-$, which is more abundant, although the higher level of background requires additional care in the analysis. It is expected that these high precision measurements will allow to put tight constraints on NP in the up-quark sector.

\end{document}